\documentclass[9pt,twocolumn,technote,twoside]{IEEEtran}

\begin{document}
\title{Limit Cycle Oscillations in Pacemaker Cells.}  \author{Lars
Petter Endresen\thanks{Institutt for fysikk, NTNU, N-7034 Trondheim,
Norway (email: endresen@phys.ntnu.no).} and Nils
Skarland$^*$\thanks{${}^*$c/o Duggan, 25 Cornwall Gardens, London SW7
4AW, UK (email: skarland@pvv.ntnu.no).}.}

\date{\today}
\maketitle

\begin{abstract}
In recent decades, several mathematical models describing the
pacemaker activity of the rabbit sinoatrial node have been developed.
We demonstrate that it is not possible to establish the existence,
uniqueness, and stability of a limit cycle oscillation in those
models. Instead we observe an infinite number of limit cycles. We then
display numerical results from a new model, with a limit cycle that
can be reached from many different initial conditions.
\end{abstract}


\section{Introduction}
\noindent
In elementary electrostatics it is well known that the relation
between the voltage and the charge of a capacitor is
\begin{equation}
\label{eq1}
q = C v \;,
\end{equation}

\noindent
where $v$ is voltage, $C$ is capasitance and $q$ is
charge. Differentiating this equation with respect to time we obtain
\begin{equation}
\label{eq2}
\frac{dq}{dt} = C \frac{dv}{dt} \;,
\end{equation}

\noindent
where $\frac{dq}{dt}\equiv i$ is a current, and the sign of $v$ is a
matter of convention. In physiology this second relation (\ref{eq2})
is being used to describe how the membrane potential ($v$) is changed
when ions move across the cell membrane. Unfortunately we observe
that this equation also is being used in some models where one in
addition keeps track of the charge ($q$) concentrations inside and
outside the cell \cite{Wilders}. In those models voltage and charge
are believed to be independent dynamic variables: first one determines
the voltage by integrating the membrane currents, then one determines
the charge by integrating the same membrane currents.

The purpose of this article is to point out that integrating the
membrane currents {\em once} is enough. Voltage and charge cannot
simultaneously be independent dynamical variables in a model, simply
because of (\ref{eq1}).

In order to visualize the drawbacks of treating voltage and charge as
independent variables, we explore numerically the nonlinear dynamics
of two different models describing the pacemaker activity of the
rabbit sinoatrial node. The procedure is as follows:
\begin{enumerate}
\item we integrate numerically the equations of motion for a
sufficiently long time to detect a steady state, 
\item we {\em change} the initial conditions and repeat 1.
\end{enumerate}

First we display results from the Wilders {\em et al.}  model
\cite{Wilders}, a model that treats voltage and charge as independent
variables. In that model it is thus possible to select an initial
voltage and an initial charge independently. The dynamics of that
model seems peculiar. An infinity of limit cycles is observed: each
time we select new initial conditions a new limit cycle, corresponding
to a new value of the constant of motion $q-Cv$, is found. This
hampers the usefulness of the model.

Second we display results from a new model of Endresen {\em et al.}
\cite{Endresen}, where the voltage is not a dynamic variable. Here we
cannot select an initial voltage independently of the initial charge,
and only one limit cycle is observed.

\section{Existing Models}
The model of Wilders {\em et al.} \cite{Wilders} of the pacemaker
activity of the rabbit sinoatrial node serves as an excellent example
of the many models where the membrane potential is thought to be
independent of the intracellular and extracellular charge
concentrations. In that model the equation of motion for the voltage
is given by (\ref{eq2})
\begin{equation}
\label{eq3}
\frac{dv}{dt} = -\frac{1}{C} (i_{\rm b,Ca}+i_{\rm b,K}+i_{\rm
b,Na}+i_{\rm Ca,L}+i_{\rm Ca,T}+i_{\rm f}+i_{\rm K}+i_{\rm Na}+i_{\rm
NaCa}+i_{\rm NaK})\;.
\end{equation}

\noindent
There are fifteen dynamic variables in that model, the voltage $v$,
the gating variables $d_{\rm L}$, $d_{\rm T}$, $f_{\rm L}$, $f_{\rm
T}$, $x$, $y$, $h$, $m$, $p$, and the ionic concentrations $[{\rm
Ca}]_{\rm i}$, $[{\rm Ca}]_{\rm rel}$, $[{\rm Ca}]_{\rm up}$, $[{\rm
K}]_{\rm i}$, $[{\rm Na}]_{\rm i}$. We want to determine how the long
term dynamics in that model is changed when we change the initial
conditions. To keep matters simple, we only change one initial
condition: the initial intracellular concentration of potassium
($[{\rm K}]_{\rm i}$); and we study the dynamics in two dimensions
only: the phase space of $v$ and $[{\rm K}]_{\rm i}$. 
\vspace*{6.8cm}
\begin{figure}[hbt]
\begin{center}
\includegraphics{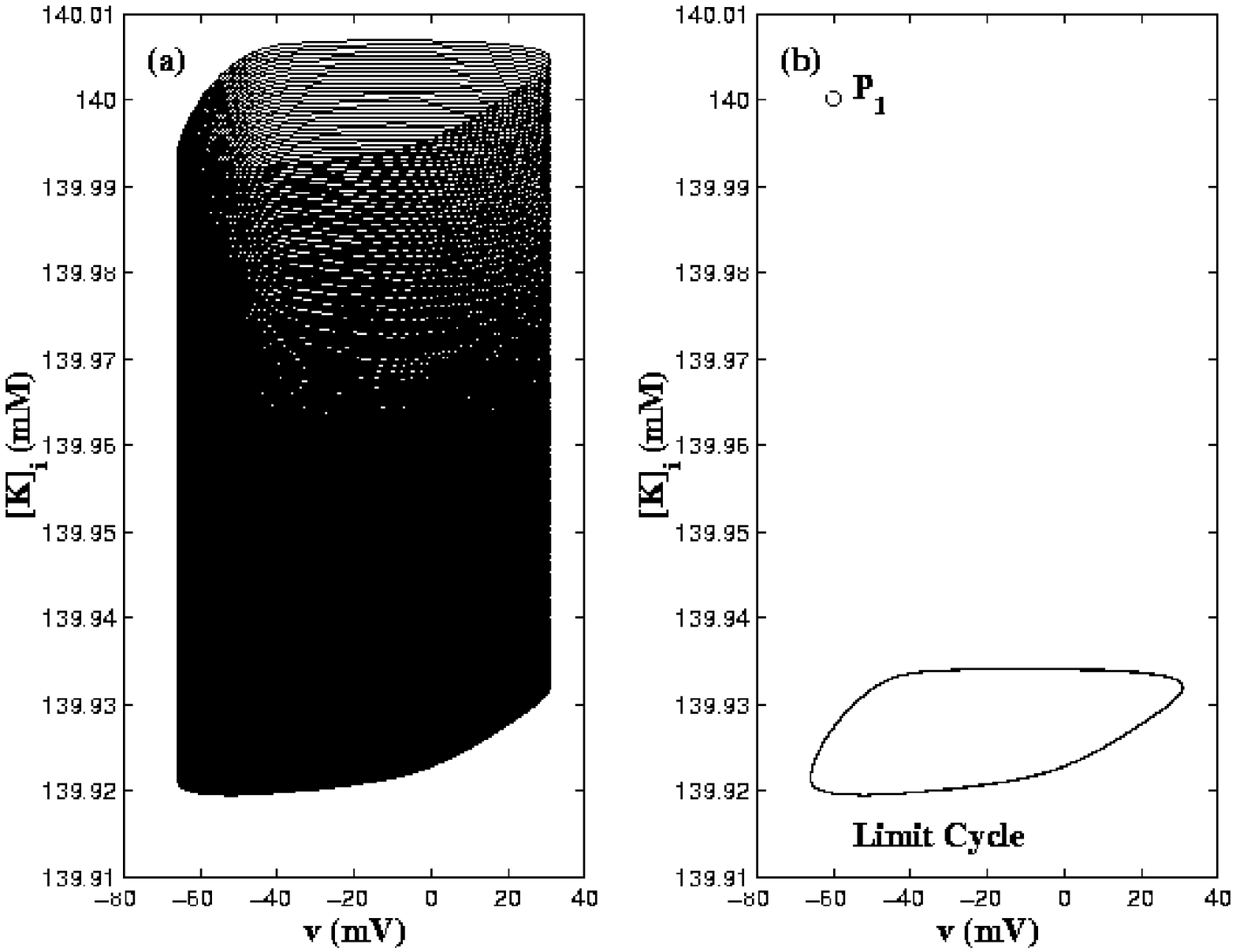}
\caption{\footnotesize Phase portrait of the Wilders {\em et al.}
model. In (a) the trajectories of the oscillator and in (b) the limit
cycle and the initial conditions (${\rm P}_1$) given by (\ref{eq6})
and ${[{\rm K}]}_{\rm i} = 140$.}
\end{center}
\end{figure}

\noindent
Figure 1 displays the two--dimensional dynamics of the model with the
initial conditions (dimensions skipped)
\begin{equation}
\label{eq6}
\begin{array}{lll} 
  v  = -60.03 \nonumber &  x  = 0.3294906 &  {[{\rm Ca}]}_{\rm i} = 0.0000804 \\
  d_{\rm L}  = 0.0002914 & y  = 0.1135163 &  {[{\rm Ca}]}_{\rm rel} = 0.6093 \\
  d_{\rm T}  = 0.0021997 &  h  = 0.1608417 &  {[{\rm Ca}]}_{\rm up} = 3.7916 \\
  f_{\rm L}  = 0.9973118 &  m  = 0.1025395 &  {[{\rm K}]}_{\rm i} = {\rm variable} \\
  f_{\rm T} = 0.1175934 &  p  = 0.2844889 &  {[{\rm Na}]}_{\rm i} = 7.5  \;,
\end{array} 
\end{equation} 

\noindent
and ${[{\rm K}]}_{\rm i} = 140$. In figure 1 (b) the point ${\rm P}_1$
denotes the $v$ and $[{\rm K}]_{\rm i}$ coordinates of the initial
conditions, and the closed loop at the bottom is the limit cycle. The
trajectory of the model is displayed in 1 (a) where we observe that
the model spirals downwards from the point ${\rm P}_1$ to the limit
cycle.
\vspace*{6.8cm}
\begin{figure}[hbt]
\begin{center}
\includegraphics{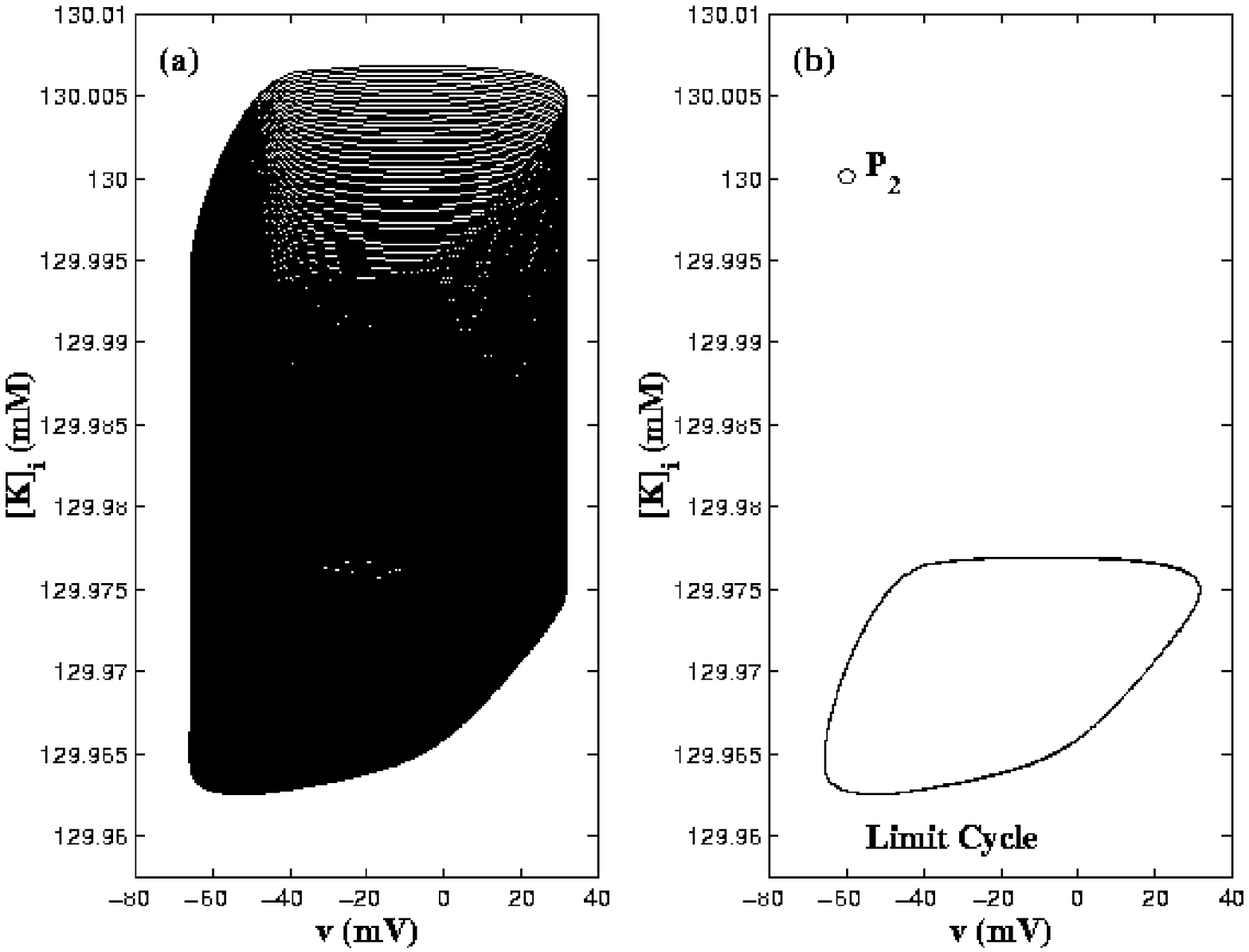}
\caption{\footnotesize Phase portrait of the Wilders {\em et al.}
model. In (a) the trajectories of the oscillator and in (b) the limit
cycle and the initial conditions (${\rm P}_1$) given by (\ref{eq6})
and ${[{\rm K}]}_{\rm i} = 130$.}
\end{center}
\end{figure}

If the limit cycle in figure 1 (b) is unique it should be possible to
reach it from another initial condition. Let us try an initial
condition below the limit cycle, and investigate whether the model
spiral up towards it. We change the initial concentration of potassium
from $140$ to $130$, leaving the fourteen other initial conditions
unchanged. The result is displayed in figure 2. The model does not
spiral upwards to the limit cycle in figure 1, instead the model
spiral downwards to a different limit cycle. We observed numerically a
new limit cycle for each new initial value of $[{\rm K}]_{\rm i}$,
implying the existence of an infinite number of limit cycles. The
model's fundamental flaw is clearly demonstrated.

\section*{A New Model}
\noindent
In a new model \cite{Endresen} of the pacemaker activity of the rabbit
sinoatrial node, the membrane potential is determined by (\ref{eq1})
\begin{equation}
\label{eq7}
v = \frac{FV}{C} \left\{[\rm K]_{\rm i}-[\rm K]_{\rm e} + 2([\rm Ca]_{\rm i}-[\rm Ca]_{\rm e}) + [\rm Na]_{\rm i}-[\rm Na]_{\rm e} \right\} \;,
\end{equation}

\noindent 
where $q=FV\left\{[\rm K]_{\rm i}-[\rm K]_{\rm e} + 2([\rm Ca]_{\rm
i}-[\rm Ca]_{\rm e}) + [\rm Na]_{\rm i}-[\rm Na]_{\rm e} \right\}$ is
the charge difference, $F$ is Faraday's constant and $V$ is cell
volume. Here the ionic currents alter the concentrations which in turn
alter the voltage, i.e. the physical quantities were calculated in the
following order: ${\rm current \;} i \Rightarrow {\rm charge \;} q
\Rightarrow {\rm voltage \;} v $. The model has five dynamic
variables, the gating variables $x$, $h$ and the ionic concentrations
$[{\rm Ca}]_{\rm i}$, $[{\rm K}]_{\rm i}$, $[{\rm Na}]_{\rm i}$, and
we use the initial conditions:
\begin{equation}
\label{eq8}
\begin{array}{l}
  x = 0.9165 \\
  h = 0.0000 \\
  {[{\rm K}]}_{\rm i} = {\rm variable} \\
  {[{\rm Ca}]}_{\rm i} = 0.004094141 \\
  {[{\rm Na}]}_{\rm i} = 18.73322695 \;.
\end{array} 
\end{equation}

\noindent
In this model we first notice that the initial value of $[{\rm
K}]_{\rm i}$, due to (\ref{eq7}), is not independent of the initial
value of the voltage $v$. Thus changing $[{\rm K}]_{\rm i}$ changes
$v$ as is always the case when charging a capacitor
(\ref{eq1}). Second we notice that a tiny change in $[{\rm K}]_{\rm
i}$ corresponds to a large change in voltage $v$, since the constant
$FV/C$ is large in most situations. 
\vspace*{6.8cm}
\begin{figure}[hbt]
\begin{center}
\includegraphics{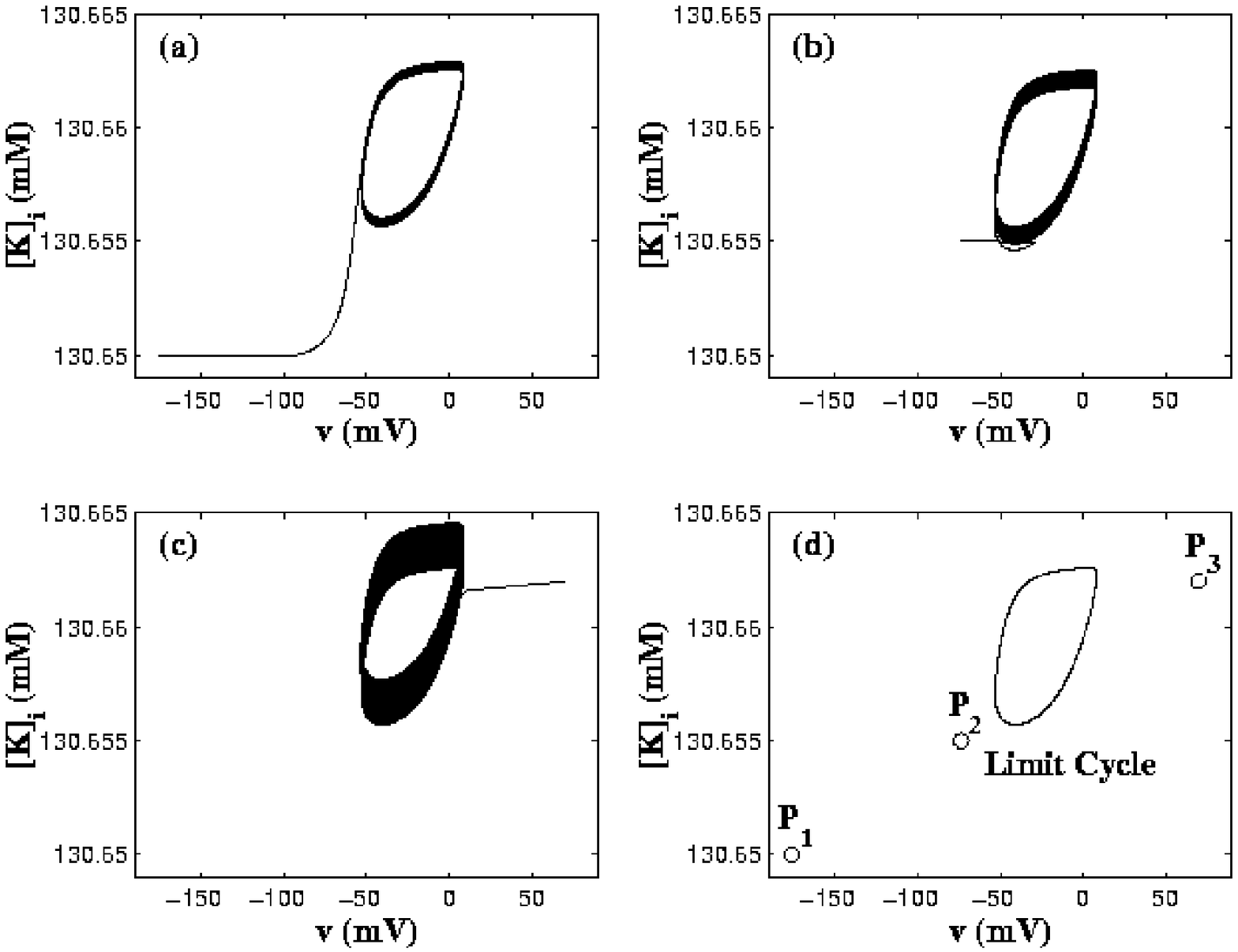}
\caption{\footnotesize Phase portrait of the Endresen {\em et al.}
model with the initial conditions (\ref{eq8}). In (a) the trajectory
with ${[{\rm K}]}_{\rm i}= 130.650$ (${\rm P}_1$), in (b) the
trajectory with ${[{\rm K}]}_{\rm i} = 130.655$ (${\rm P}_2$), in (c)
the trajectory with ${[{\rm K}]}_{\rm i} = 130.662$ (${\rm P}_3$), and
in (d) the unique limit cycle and the three initial conditions ${\rm
P}_1$, ${\rm P}_2$ and ${\rm P}_3$ from (a), (b) and (c).}
\end{center}
\end{figure}

In figure 3 we have displayed the simulation results from the model of
Endresen {\em et al.} \cite{Endresen} with three slightly (due to the
large constant $FV/C$) different initial values of ${[{\rm K}]}_{\rm
i}$: 130.650 (a), 130.655 (b), and 130.662 (c). In figure 3 (d) the
three initial conditions ${\rm P}_1$, ${\rm P}_2$ and ${\rm P}_3$ all
converge towards the same limit cycle. In an extensive numerical study
we have not observed any physiological initial conditions that do not
converge toward this limit cycle. In fact the same limit cycle can be
reached when starting from the full equilibrium situation with equal
intracellular and extracellular ionic concentrations \cite{Endresen}.

\section{Discussion}
\noindent
We have displayed numerical results from two types of mathematical
models of the pacemaker activity of the rabbit sinoatrial node. The
first type of model \cite{Wilders} showed an infinite number of limit
cycles, the second type of model \cite{Endresen} a limit cycle that
could be reached from many different initial conditions. In order to
avoid the drawback with an infinite number of limit cycles seen in the
first type of models, we suggest that one should not treat membrane
voltage ($v$) as a dynamic variable. Instead one should calculate the
voltage using (\ref{eq1}), or at least select the initial conditions
in agreement with (\ref{eq1}) \cite{Endresen}.

\section*{Acknowledgments}
Lars Petter Endresen was supported by a fellowship at NTNU, and has
received support from The Research Council of Norway (Programme for
Supercomputing) through a grant of computing time.

\nocite{*}
\bibliographystyle{IEEE}

\begin{thebibliography}{1}
\bibitem{Endresen} \newblock L.P. Endresen, K. Hall, J.S. H{\o}ye, and
J. Myrheim, ``A Theory for the Membrane Potential of Living Cells,''
{\em European Biophysics Journal,} Submitted, 1999.

\bibitem{Wilders} \newblock R. Wilders, H.J. Jongsma, and A.C. van
Ginneken, ``Pacemaker activity of the rabbit sinoatrial node. A
comparison of mathematical models,'' {\em Biophysical Journal,}
vol. 60, pp. 1202--1216, 1991.
\end{thebibliography}

\end{document}